\documentclass[dvips]{article}

\usepackage{icrc2011}

\title{Highlights of the VERITAS Blazar Observation Program}
\shorttitle{W. Benbow:  VERITAS Blazar Program}

\authors{Wystan Benbow$^{1}$ for the VERITAS Collaboration$^{2}$ }
\afiliations{$^1$Harvard-Smithsonian Center for Astrophysics, 60 Garden St., MS-20, Cambridge, MA, 02180, USA\\$^2$see J. Holder et al. (these proceedings) or http://veritas.sao.arizona.edu/conferences/authors?icrc2011}
\email{wbenbow@cfa.harvard.edu}

\abstract{
The VERITAS array of 12-m atmospheric-Cherenkov telescopes 
in southern Arizona began full-scale operations in 2007, and 
it is one of the world's most sensitive detectors of astrophysical 
VHE ($E>$100 GeV) $\gamma$-rays.  Forty-one blazars are 
known to emit VHE photons, and
observations of blazars are one of the VERITAS Collaboration's 
Key Science Projects (KSPs). More than 400 hours per year are devoted
to this program, and $\sim$100 blazars have already been observed 
with the array, in most cases with the deepest-ever VHE exposure.
These observations have resulted in 20 detections, 
including 10 new VHE blazars.  Highlights of the VERITAS 
blazar observation program, and the collaboration's long-term blazar 
observation strategy, are presented.}
\keywords{VERITAS, AGN, Blazar, Gamma-ray, TeV, VHE}

\begin{document}
\maketitle

%Begin the section.

\vspace{-0.2cm}
\section{Introduction}
\vspace{-0.2cm}
For almost two decades, the study of 
blazars has been a major component of the scientific
program of VHE $\gamma$-ray observatories.
Forty-one blazars, a class of AGN with relativistic jets pointed along 
the line of sight to the observer, are observed to emit VHE $\gamma$-rays.
The VHE blazar population includes four blazar subclasses:
30 high-frequency-peaked BL\,Lac objects (HBLs),
5 intermediate-frequency-peaked BL Lac objects (IBLs), 
3 low-frequency-peaked BL Lac objects (LBLs), and
3 flat-spectrum radio quasars (FSRQs).  The redshifts
of the known VHE blazars range from $z = 0.030$ to at
least $z = 0.536$, and the photon spectra of the
observed VHE emission is often soft 
($\Gamma_{obs} \sim 2.5 - 4.6$).  However, it
is important to note that this is largely due
to the softening of the emitted blazar spectra by the
attenuation of VHE photons on the extragalactic background
light (EBL).  Approximately 60\% of the known HBLs
exhibit some variability, and most of 
the non-HBL blazars are detected 
at VHE only during flaring episodes.
Although VHE flux variability is commonly observed\footnote{Typically
variations of a factor of 2-3 are seen on timescales ranging from days to years.},
rapid (minute-scale), large-scale (factor of 100) variations 
of the VHE flux are relatively rare (see, e.g., \cite{PKS2155_flare}).  
Only four VHE blazars
have ever been observed at more than the Crab Nebula flux (1 Crab),
and of these, only Mkn\,421 and Mkn\,501 have
been seen in multiple $>$1 Crab flaring episodes.

Understanding VHE blazars and their related science relies
on expanding the known population, and making precision
measurements of their spectra (particularly at the highest energies)
and their variability patterns (e.g., timescales, flux range, 
and spectral changes).  Contemporaneous multiwavelength (MWL) 
observations are a key component of these studies since these 
highly-variable sources emit
over the entire broadband spectrum.  These MWL studies
enable modeling of the blazars' double-humped 
spectral energy distributions (SEDs), as well as searches for
correlations in the flux/spectral changes observed that
may indicate commonalities in the origin of the observed emission.

\vspace{-0.3cm}
\section{The VERITAS Blazar KSP}
\vspace{-0.2cm}

\begin{table}[t]
\begin{center}
\begin{tabular}{c | c | c | c }
\hline

{\footnotesize Blazar} & {\footnotesize $z$} &  {\footnotesize Type} & {\footnotesize log$_{10}(\nu_{\rm synch})$}\\
\hline
{\footnotesize Mrk\,421} & {\footnotesize 0.030} & {\footnotesize HBL} & {\footnotesize 18.5}\\
{\footnotesize Mrk\,501} & {\footnotesize 0.034} & {\footnotesize HBL} & {\footnotesize 16.8}\\
{\footnotesize 1ES\,2344+514} & {\footnotesize 0.044} & {\footnotesize HBL$^{\beta}$} & {\footnotesize 16.4}\\
{\footnotesize 1ES\,1959+650} & {\footnotesize 0.047} & {\footnotesize HBL} & {\footnotesize 18.0}\\
{\footnotesize W\,Comae$^{\dagger}$} & {\footnotesize 0.102} & {\footnotesize IBL} & {\footnotesize 14.8}\\
{\footnotesize RGB\,J0710+591$^{\dagger}$} & {\footnotesize 0.125} & {\footnotesize HBL} & {\footnotesize 21.1}\\
{\footnotesize H\,1426+428} & {\footnotesize 0.129} & {\footnotesize HBL} & {\footnotesize 18.6}\\
{\footnotesize 1ES\,0806+524$^{\dagger}$} & {\footnotesize 0.138} & {\footnotesize HBL} & {\footnotesize 16.6}\\
{\footnotesize 1ES\,0229+200} & {\footnotesize 0.140} & {\footnotesize HBL} & {\footnotesize 19.5}\\
{\footnotesize 1ES\,1440+122$^{\dagger}$} & {\footnotesize 0.162} & {\footnotesize IBL} & {\footnotesize 16.5}\\
{\footnotesize RX\,J0648.7+1516$^{\dagger}$} & {\footnotesize 0.179} & {\footnotesize HBL$^{*}$} & {\footnotesize - }\\
{\footnotesize 1ES\,1218+304} & {\footnotesize 0.184} & {\footnotesize HBL} & {\footnotesize 19.1 }\\
{\footnotesize RBS\,0413$^{\dagger}$} & {\footnotesize 0.190} & {\footnotesize HBL} & {\footnotesize 17.0}\\
{\footnotesize 1ES\,0414+009} & {\footnotesize 0.287} & {\footnotesize HBL} & {\footnotesize 20.7}\\
{\footnotesize PG\,1553+113} & {\footnotesize $0.43 < z < 0.50$} & {\footnotesize HBL$^{\beta}$} & {\footnotesize 16.5}\\
{\footnotesize 3C\,66A$^{\dagger}$} & {\footnotesize ?} & {\footnotesize IBL} & {\footnotesize 15.6}\\
{\footnotesize B2\,1215+30} & {\footnotesize ?} & {\footnotesize IBL} & {\footnotesize 15.6}\\
{\footnotesize PKS\,1424+240$^{\dagger}$} & {\footnotesize ?} & {\footnotesize IBL} & {\footnotesize 15.7}\\
{\footnotesize 1ES\,0502+675$^{\dagger}$} & {\footnotesize ?} & {\footnotesize HBL} & {\footnotesize 19.2}\\
{\footnotesize RGB\,J0521.8+2112$^{\dagger}$} & {\footnotesize ?} & {\footnotesize HBL$^{*}$} & {\footnotesize -}\\
\hline
\end{tabular}
\vspace{-0.2cm}
\caption{{\footnotesize The 20 blazars detected at VHE with VERITAS. The 10
VHE discoveries are marked with $\dagger$. 
The catalog redshifts for B2\,1215+30 (0.130), 1ES\,0502+675 (0.341) 
and 3C\,66A (0.444) are considered uncertain, and the redshift
range for PG\,1553+113 is quoted at the 
lower and upper limits determined by \cite{1553_z_HST} and \cite{1553_ICRC},
respectively. The classifications
and synchrotron peak frequencies are taken from \cite{Nieppola}, 
except for four cases: two (marked with asterisks) where the
classification is determined from VERITAS-led MWL studies,
and two (marked with $\beta$) where the
historical HBL classification is used.}}\label{blazar_table}
\end{center}
\vspace{-0.3cm}
\end{table}

\begin{figure}[t]
  \vspace{5mm}
  \centering
  \includegraphics[width=2.in]{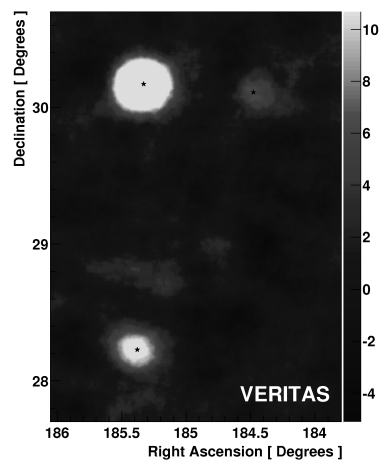}
\vspace{-0.2cm}
\caption{{\footnotesize The significance map for a single
field of view containing three VHE blazars (1ES\,1218+304, 
W\,Comae \& B2\,1215+30).
All three sources are point-like, but appear to have
a different size due to the saturation on the color scale.
The total exposure for this map is $\sim$130 h of observations,
but the effective exposure at each blazar position varies
considerably because the three sources were initially observed
individually.}}
  \label{3src_map}
\vspace{-0.3cm}
 \end{figure}

VERITAS began routine scientific observations with the full array in 
September 2007. The performance metrics of VERITAS \cite{Holder_ICRC} 
include an energy resolution of $\sim$15\%, an angular resolution 
of $\sim$0.1$^{\circ}$, and a sensitivity yielding
a 5 standard deviation ($\sigma$) 
detection of an object with flux equal to
1\% Crab in $\sim$25 hours. 
VERITAS observations are performed for $\sim$1100 h 
each year, and from 2007-10, observations of blazars averaged 
$\sim$410 h per year.  Table~\ref{blazar_table} shows
the 20 VHE blazars (15 HBL and all 5 
known VHE IBL) detected by VERITAS, which include
10 discoveries.  The VERITAS blazar KSP consists of:

{\bf i. A discovery program:} Several
blazars are selected annually for VERITAS observations with the goal
of increasing the VHE blazar population, particularly in
the non-HBL subclasses and at higher redshifts.  Through these 
efforts, and those of other VHE observatories, the VHE blazar catalog 
is rapidly expanding.  This is well illustrated by the sky map
in Figure~\ref{3src_map}, which shows 3 VHE blazars
(2 IBLs and 1 HBL) all contained within the VERITAS field
of view (3.5$^{\circ}$). Highlights from the
discovery program are presented elsewhere in
these proceedings \cite{Benbow_ICRC11_disc}.

{\bf ii. A target-of-opportunity (ToO) observation program:}
Blazar observations can be triggered by 
either a VHE discovery or a flaring alert from various optical,
X-ray, MeV-GeV (Fermi-LAT), and TeV blazar monitoring programs.

{\bf iii. A MWL observation program:} Contemporaneous MWL
observations are organized for
most of the known VHE blazars regularly observed by VERITAS.  In addition,
ToO observation proposals for MWL measurements are also submitted to 
lower-energy observatories, and are triggered by a VERITAS 
discovery or flaring alert.

{\bf iv. An EBL/IGMF program:} 
Studies of distant VHE candidates, and deep exposures on known
hard-spectrum VHE blazars, are performed to constrain, 
and possibly measure, the EBL
and the intergalactic magnetic field (IGMF).

From 2007-10,
$\sim$80\% of the VERITAS blazar data came from discovery 
observations and follow-up observations of any new sources.  
In September 2010, the emphasis of the Blazar KSP changed
to aim for a 40:60 ratio between discovery observations
and exposures of known VHE sources. 

\vspace{-0.3cm}
\section{Highlights for Known VHE Blazars}
\vspace{-0.2cm}
{\bf Mrk 421} is the longest-known VHE blazar, and
generally has the brightest VHE flux.
It is easily the best-studied HBL at VHE, and 
VERITAS has acquired nearly 80 h on this blazar 
since 2007, largely during flaring states identified
with the Whipple 10-m telescope.
A total of 47 hrs of VERITAS and 96 hrs of Whipple 10m data 
taken between 2006 and 2008 are presented in 
\cite{Mkn421_beilicke}.  During this campaign, bright VHE
flares reaching flux levels of $\sim$10 Crab are
detected, and the VHE data are complemented with 
radio, optical, and X-ray (RXTE and Swift) observations. 
Flux variability is found in all bands except for the radio waveband. 
Interestingly, the VHE and X-ray flux are often correlated,
with both bands showing spectral hardening with increased flux levels.
For 18 nights during a 118-day campaign in 2008, 
it was possible to generate contemporaneous SEDs, 
each of which can be described by a 
one-zone synchrotron-self-Compton (SSC) model.
VERITAS monitoring of the VHE flux from Mrk\,421 in 2009-10 reveals an
elevated state during the entire season.  In particular,
an extreme flare was observed for nearly 5 h live time on February 17, 2010,
during which the VHE flux averaged $\sim$8 Crab and showed variations
on timescales of approximately 5-10 minutes \cite{Galante_ICRC}.
Results from the extensive
MWL observations during the 2009-10 season,
including $\sim$20 h of VERITAS data, are in preparation.
During the 2010-11 season, strong
variations of the VHE flux from Mrk\,421 were observed, although, 
the flux never exceeded 1 Crab.

{\bf Mrk\,501} is perhaps the best-studied VHE blazar
after Mrk\,421.  Its VHE emission is relatively bright, and it 
has a long history of extreme outbursts
and spectral variability.  The VERITAS collaboration has performed
several MWL observation campaigns on this HBL:
during low states in May-June 2008 \cite{Mkn501_2008} and
March 2009 \cite{Mkn501_low}, and during/after a VHE flare initially
detected by the Whipple 10-m \cite{Pichel_501} in 
late-April 2009 \cite{Mkn501_LAT}.  The low-state SED measured
during the March 2009 campaign (including
Suzaku, Fermi-LAT, MAGIC and VERITAS data), contrasts remarkably
with the SED from an extreme outburst observed on April 16, 1997.
Here the energy of the X-ray peak differs by over two orders of magnitude 
between the two states, while the VHE peak location
varies little, likely due to the onset of Klein-Nishina effects. 
Regardless of this, a simple SSC scenario can successfully model
both states with the primary difference being the injected electron
spectra.  The entire 2009 VERITAS data set is presented 
in \cite{Mkn501_LAT} as part of 
an unprecedented campaign with instruments including Fermi-LAT, MAGIC,
Swift,  VLBA,  RXTE, F-GAMMA, GASP-WEBT, and other smaller
observatories.  The measured SED is the most detailed ever, and is again
well described by a standard one-zone SSC model.  In 2011, strong variations
of the VHE flux from Mrk\,501 were observed, but the VHE flux never
exceeded 1.5 Crab.

{\bf 1ES 2344+514} is the third blazar
detected at VHE, and this HBL was observed by
VERITAS for 18.1 h quality-selected live time
as part of an intense MWL observation campaign from
October 2007 to January 2008.  The VERITAS observations
yield a strong (20$\sigma$) detection of a variable
VHE flux \cite{2344_paper}.  On December 7, 2007, a strong VHE flare of 48\% Crab
above 300 GeV was observed by VERITAS.  Excluding this flare, the measured
VHE flux is still variable and averages 7.6\% Crab.  The
VHE spectrum on the night of the flare ($\Gamma = 2.43 \pm 0.22$)
is harder, but does not differ significantly from that
determined with VERITAS from the rest of the data ($\Gamma = 2.78 \pm 0.09$).
Both the VHE flux and the X-ray flux (Swift \& RXTE) vary by
a factor of $\sim$7 during the campaign, and significant
correlations between the two bands are found.  A one-zone
SSC model can describe the SEDs determined both
during the flare and in the lower-flux state.  VERITAS
monitoring of the VHE flux from 1ES\,2344+514 in 2010 yielded 
the lowest value ($\sim$2\% Crab) ever recorded from this object.

{\bf 1ES\,1959+650}, a well-studied HBL,
was observed by VERITAS for 5.3 h of quality-selected
live time in 2007-09.  Analysis of these
data yields an excess of $\sim$150 $\gamma$-rays (12.3$\sigma$),
corresponding to a flux of $\sim$18\% Crab.  The observed
flux and photon spectrum are consistent with those measured
during previous low-emission states of this source.  A MWL
observation campaign on 1ES\,1959+650, with VERITAS, Fermi-LAT and RXTE,
is planned in 2011.

{\bf H\,1426+428} was first detected 
during an outburst in 2001 \cite{Whipple_1426}.  
This HBL was observed by VERITAS for $\sim$22 h quality-selected live time
between 2007 and 2011.  A weak excess, 5.2$\sigma$,
is observed in these data, marking the
first time H\,1426+428 is detected since 2002.
The observed flux is $<$2\% Crab,
well below the value (13\% Crab) reported during
its VHE discovery, and also below any other published VHE flux 
or limit from this source.

{\bf 1ES\,1218+304} is an HBL with a VHE spectrum that is
unusually hard considering its redshift and the 
effect of the EBL.  VERITAS first detected
this blazar during 17.4 h of commissioning-phase
observations in January - March 2007 \cite{1218_paper1}.  
The $\sim$10$\sigma$ detection
corresponds to an observed flux of $\sim$6\% Crab above 200 GeV,
and the resulting spectrum between 160 GeV and
1.8 TeV is well described by a power-law function
with photon index $\Gamma=3.08\pm0.34$.  In 2008-09,
VERITAS monitored the flux from 1ES\,1218+304 for $\sim$27 h 
good-quality live time.  The blazar is strongly detected ($\sim$22$\sigma$), 
and clear day-scale variations of the VHE flux are seen \cite{1218_paper2}.  
Although the VHE flux varied by factor of $\sim$4, reaching
$\sim$20\% Crab, the VHE spectrum did not change significantly during the
campaign ($\Gamma_{avg} =3.07\pm 0.09$).  The relative hardness of the
1ES\,1218+304 spectrum can be used to derive 
limits on the EBL very close to the best. However,
the derived EBL constraints can be weakened by invoking
kpc-scale jet-emission scenarios for this
and other distant, hard-spectrum VHE blazars.
Fortunately, the observed day-scale
variability rules these models out \cite{1218_paper2}.
During VERITAS monitoring of 1ES\,1218+304 in 2011, the blazar
is strongly detected in a season-long elevated flux state ($\sim$12\% Crab).

{\bf PG\,1553+113} is a hard-spectrum ($\Gamma_{LAT} \sim 1.66$)
Fermi-LAT blazar  \cite{Fermi_1FGL}.
It is likely the most distant HBL detected at VHE (see $z > 0.43$ from
\cite{1553_z_HST}). It was observed by VERITAS for $\sim$60 h of quality-selected
live time between May 2010 and May 2011.  These data
result in the most significant VHE detection (39$\sigma$)
of this HBL.  The time-averaged VHE flux is 10\% Crab above
200 GeV, higher than the archival VHE measurements, and
the photon spectrum is well described
between 175 GeV and 500 GeV by a power-law function
with photon index $\Gamma = 4.41 \pm 0.14$.
The VHE spectrum can be used to set an upper limit on the redshift
of $z < 0.5$. More details regarding the VERITAS detection of PG\,1553+113
can be found in these proceedings \cite{1553_ICRC}.

{\bf 1ES\,0229+200} is one of the hardest-spectrum VHE
blazars known ($\Gamma_{HESS} \sim 2.5$; \cite{HESS_0229}).  
It was observed by VERITAS as part of an intense MWL observation campaign
for $\sim$46 h live time from 2009-11.  
A strong signal is detected ($\sim$600 $\gamma$-rays,
9.0$\sigma$) in these observations corresponding to an average VHE flux
of $\sim$2\% Crab above 300 GeV.  The VERITAS flux is variable
on a timescale of months, and the preliminary VHE spectrum
measured between $\sim$220 GeV and $\sim$15 TeV 
has photon index $\Gamma = 2.44 \pm 0.11$.  The results of the
MWL campaign are in preparation. It is interesting to note that
1ES\,0229+200 is the only VERITAS-detected blazar not
included in the 1FGL catalog \cite{Fermi_1FGL}.

{\bf 1ES\,0414+009}
1ES\,0414+009 is the most distant VHE HBL with a
well-measured redshift ($z = 0.287$).  It was 
observed by VERITAS for $\sim$55 h of quality-selected
live time from January 2008 to February 2011.  
An excess of VHE $\gamma$-rays is detected ($\sim$7$\sigma$)
from this Fermi-LAT source ($\Gamma_{LAT} \sim 1.94$; \cite{Fermi_1FGL}).
The observed VERITAS spectrum between $\sim$230 GeV and
$\sim$1.8 TeV is relatively hard ($\Gamma = 3.4 \pm 0.5$)
considering EBL-related effects, and consistent
with that observed during the HESS discovery \cite{HESS_0414}.  
The observed VERITAS flux is somewhat higher ($\sim$1.6\% Crab) 
than measured by HESS (0.6\% Crab above 200 GeV), although the 
large data-sets used by both experiments are not simultaneous.
Results from a contemporaneous MWL observation campaign are
in preparation.

{\bf B2\,1215+30}, an IBL discovered
at VHE during a flare in January 2011\cite{MAGIC_1215},
was observed\footnote{Almost all
of these data were taken as 0.5$^{\circ}$ wobbles
on another VHE blazar, 1ES\,1218+304, $\sim$0.9$^{\circ}$
distant, resulting in a lower average sensitivity for the VERITAS exposure.}
for $\sim$55 h of quality-selected live time between December 2008 and April
2011.  The measured excess of $\sim$240 $\gamma$-rays (6.3$\sigma$)
corresponds to a VHE flux of $\sim$1\% Crab.  There
is a weak indication that the flux 
observed by VERITAS in 2011 may be higher than seen from 2008-10.  
The VERITAS flux is consistent 
with that ($2\pm1$\% Crab) reported during the MAGIC discovery.

\vspace{-0.3cm}
\section{The Long-term Blazar Strategy}
\vspace{-0.2cm}
In 2010, a plan for the future of the 
VERITAS blazar KSP was developed.
Using the average annual blazar exposure, $\sim$410 h per year,
as the baseline, an observation program was created that
focused on intensive observations of known sources, while
maintaining an active discovery effort. 
The major components of this program include:

{\bf i. Long-term Monitoring (210 h / yr):}  VERITAS will 
regularly monitor 14 selected VHE blazars ($\sim$70\% of northern 
VHE BL Lac population) during each season to maximize the 
chance of detecting any VHE flaring episodes 
while simultaneously building deep exposures ($\sim$100 to $\sim$200 h total). 
The selected targets consist of five EBL/IGMF-relevant HBLs
(i.e., distant, hard-spectrum sources), 
four nearby, bright HBL where extreme flares are perhaps most likely, 
and five non-HBL blazars for studies to 
unravel the mechanisms behind the blazar sequence.  Table~\ref{LTP_table}
shows the blazars monitored and their existing VERITAS exposure.
Contemporaneous radio, optical/UV, X-ray and GeV monitoring 
will be organized to enable source modeling, 
and ToO observation proposals in the optical-to-X-ray waveband 
will be submitted annually to ensure coverage of important flaring events.

{\bf ii. Discovery Program ($\sim$100 h / yr):} A Fermi-LAT-guided discovery 
effort will be continued, but the focus will move towards 
higher-risk/higher-reward endeavors, and will be split between 
non-HBL targets (to expand the understanding of the blazar sequence), 
and high-redshift candidates (useful for EBL/IGMF studies).  
In addition, observations of previously-viewed candidates showing
a marginal excess in existing VERITAS data will continue.

{\bf iii. ToO Program ($\sim$100 h / yr):} This is used to 
respond to flares from the monitoring program, to deepen 
exposures on new discoveries, and to respond to high-value 
discovery opportunities (e.g., FSRQs) indicated by flaring at lower energy.  
To aid this effort, optical, and often X-ray monitoring, of all 
known VHE blazars and high-value candidates is 
set up, in addition to automatic LAT 
analysis pipelines\cite{Errando_Orr_ICRC}.

\begin{table}[t]
\begin{center}
\begin{tabular}{c | c | c }
\hline
{\footnotesize VHE} & {\footnotesize Primary} & {\footnotesize  Exp.} \\ 
{\footnotesize Blazar} & {\footnotesize Study} & {\footnotesize [h]} \\
\hline
{\footnotesize 1ES\,0229+200} & {\footnotesize EBL/IGMF } & {\footnotesize 51}\\
{\footnotesize 1ES\,0414+009} & {\footnotesize EBL/IGMF } & {\footnotesize 55}\\
{\footnotesize RGB\,J0710+591} & {\footnotesize EBL/IGMF} & {\footnotesize 46}\\
{\footnotesize 1ES\,1218+304} & {\footnotesize EBL/IGMF } & {\footnotesize 81}\\
{\footnotesize PG\,1553+113} & {\footnotesize EBL/IGMF  } & {\footnotesize 65}\\
{\footnotesize Mrk\,421} & {\footnotesize Bright HBL} & {\footnotesize 78 }\\
{\footnotesize Mrk\,501} & {\footnotesize Bright HBL} & {\footnotesize 26}\\
{\footnotesize 1ES\,1959+650} & {\footnotesize Bright HBL} & {\footnotesize 5}\\
{\footnotesize 1ES\,2344+514} & {\footnotesize Bright HBL} & {\footnotesize 24 }\\
{\footnotesize 3C\,66A} & {\footnotesize IBL} & {\footnotesize 44}\\
{\footnotesize W\,Comae} & {\footnotesize IBL} & {\footnotesize 63}\\
{\footnotesize PKS\,1424+240} & {\footnotesize IBL} & {\footnotesize 45}\\
{\footnotesize S5\,0716+714} & {\footnotesize LBL} & {\footnotesize 13}\\
{\footnotesize BL\,Lac} & {\footnotesize LBL} & {\footnotesize 13}\\
\hline
\end{tabular}
\vspace{-0.2cm}
\caption{{\footnotesize The 14 blazars selected for long-term
monitoring with VERITAS, and their existing
 good-quality VERITAS exposure (as of May 31, 2011). 
The exposure goal for each target, which may be exceeded or revised,
ranges from 5 h to 25 h per year. }}\label{LTP_table}
\vspace{-0.6cm}
\end{center}
\end{table}

\vspace{-0.3cm}
\section{Conclusion}
\vspace{-0.2cm}
Blazar observations are a major component ($\sim$40\%) of the 
scientific program of VERITAS.
Twenty VHE blazars are detected with the observatory, including
10 VHE discoveries.  The VERITAS studies of these emitters
are largely the most sensitive ever, and have yielded a number of
interesting results.  A long-term plan for
the VERITAS blazar KSP is organized, and it includes both a robust
discovery effort, and a program to produce unprecedented 
VHE data sets, in terms of sensitivity, duration and MWL coverage, 
for most of the northern blazar population.  Clearly, there are
many exciting VERITAS blazar results still to come.

\vspace{0.1cm}
{\footnotesize
This research is supported by grants from the US Department of Energy, 
the US National Science Foundation, and the Smithsonian Institution, 
by NSERC in Canada, by Science Foundation Ireland, and by STFC in the UK. 
We acknowledge the excellent work of the technical support staff at
FLWO and the collaborating institutions in the 
construction and operation of the instrument.}

\clearpage

\end{document}